# Towards edge engineering of two-dimensional layered transition-metal dichalcogenides by chemical vapor deposition

Wei Fu[1,†,*], Mark John[1,2,†], Thathsara D. Maddumapatabandi[1], Fabio Bussolotti[1], Yong Sean Yau[1], Ming Lin[1] and Kuan Eng Johnson Goh[1, 2, 3,*]

[1] Institute of Materials Research and Engineering (IMRE), Agency for Science Technology and Research (A*STAR), 2 Fusionopolis Way, Innovis #08-03, Singapore 138634, Republic of Singapore

[2] Department of Physics, National University

of Singapore, 2 Science Drive 3, Singapore 117551, Singapore

[3] Division of Physics and Applied Physics, School of Physical and Mathematical Sciences, Nanyang Technological University, 50 Nanyang Avenue 639798, Singapore

[*] Authors to whom any correspondence should be addressed.

† Authors contributed equally to this paper

E-mail: LDM.fuw@gmail.com; kejgoh@yahoo.com

**Abstract**

The manipulation of edge configurations and structures in atomically thin transition metal dichalcogenides (TMDs) for versatile functionalization has attracted intensive interest in recent years. The chemical vapor deposition (CVD) approach has shown promise for TMD edge engineering of atomic edge configurations (1H, 1T or 1T'-zigzag or armchair edges), as well as diverse edge morphologies (1D nanoribbons, 2D dendrites, 3D spirals, etc). These rich-edge TMD layers offer versatile candidates for probing the physical and chemical properties, and exploring new applications in electronics, optoelectronics, catalysis, sensing and quantum field. In this review, we present an overview of the current state-of-the-art in the manipulation of TMD atomic edges and edge-rich structures using CVD. We highlight the vast range of unique properties associated with these edge configurations and structures and provide insights into the opportunities afforded by such edge-functionalized crystals. The objective of this review is to motivate further research and development efforts in using CVD as a scalable approach to harness the benefits of such crystal-edge engineering.

## 1. Introduction

Two dimensional layered materials have attracting significant attention due to remarkable properties that set them apart from their 3D counterparts, and their flexibility for functional modulation.[1-7] Their atomically thin layered structure (limited to just a few atomic layers), provides opportunities for exploring new physics and remarkable properties.[1, 5, 8-12] The assembly and relative rotation of different 2D layered materials can lead to new phases and Moiré patterns, resulting in new physical phenomena and characteristics including strongly correlated transport behaviors.[11, 13-19] The properties of such layered materials can also be fine-tuned through strain and defect engineering.[20-23] Combined features create an important materials platform for the development of next-generation technologies in fields such as electronics[8, 24, 25], photonics[26, 27], energy[28, 29], and quantum application[30, 31]. In addition to in-plane structural engineering, the study of edge structure engineering has emerged as a vibrant area of research.[32-38] The investigation of edge structures in 2D layered materials has uncovered a wealth of atomic edge structures and edge morphologies. These edge features possess distinct properties that are not present in their planar counterparts.[37, 39, 40] This has spurred a growing body of research aimed at understanding the relationship between edge structure and the properties of 2D materials, as well as developing new methods for manipulating and controlling their edge structures for various applications. For example, as a representative 2D material, graphene nanoribbons host spin-ordered electronic states that result in quantum magnetism. These intrinsic magnetic states are derived from the zigzag edge structure of graphene nanoribbons and open up avenues for investigating carbon-based spintronics and qubits. [33, 35, 36, 38] Such properties have inspired the development of high-speed, low-power spin-logic devices for data storage and information processing.

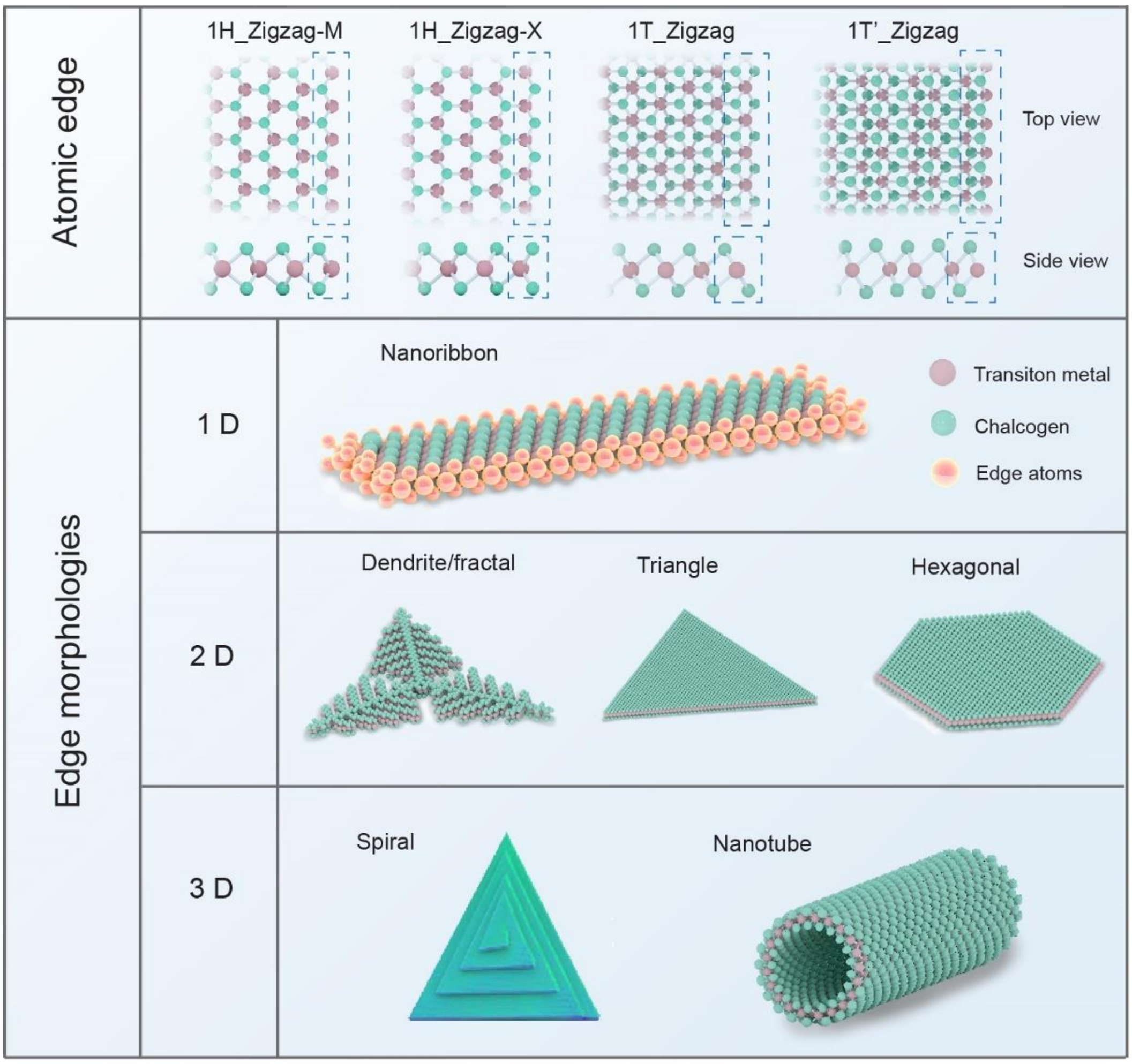

**Figure 1** Schematic illustration of typical atomic edge configuration and edge morphologies in the CVD grown TMD atomic layers

TMDs is a class of 2D materials that have attracted notable attention because of their diverse properties and natural abundance. TMDs are composed of a transition metal layer sandwiched between two chalcogenide layers with the chemical formula $MX_2$, where M stands for Mo, W, Hf, Nb, Ta, et al, and X stand for S, Se or Te.[1, 5, 41, 42] There are more than 40 different types of combinations of TMDs, ranging from insulators like $HfS_2$, semiconductor like $MoS_2$, $WS_2$, and semimetal such as $TiSe_2$, to metals such as $NbSe_2$.[43-45] Furthermore, TMDs also have three well-known structural polymorphs of 2H, 1T and 1T'. The 2H phase has a trigonal prismatic coordination pattern, whereas the coordination patterns in the 1T and 1T' phases are octahedral and distorted octahedral, respectively.[46-52] Beside the versatile chemical composition and tuneable phases, atomic TMD layers, like graphene and h-BN, also exhibit rich atomic edge configurations of zigzag and armchair atomic edges with unique electronic properties.[53-58] For example, referring to 1H-$MoS_2$, the zigzag edge is metallic, whereas the armchair edge is semiconducting[59]. Thus the regulation of these edge structures in varied phase TMDs atomic layers can result in distinct properties, including changes in electrical susceptibility, the emergence of ferromagnetic behaviour, nonlinear optical effects, enhanced catalytic efficiency, with potential utilization in the fields of spintronics, sensing, and others.[60-64]

Chemical vapor deposition (CVD) is a widely used technique for the synthesis of TMD atomic layers, along with other methods such as molecular beam epitaxy (MBE) and physical vapour deposition (PVD).[8, 65, 66] Flexibly controlling the rich edges structures of TMDs would be challenging for MBE and PVD, due to limitations posed by the high-melting-point metal source and ultra-high vacuum growth conditions required in MBE[67-70], as well as the low diffusion rate of precursor in PVD[71, 72]. CVD offers several advantages in flexible edge structure control in as-grown TMDs, like good tunability, scalability and low cost consumables.[43, 46, 73-75] The flexibility in the selection of various growth control strategies, such as substrate engineering, flow modulation, and a variety of precursor candidates, coupled with variable pressure control, renders CVD a versatile platform for the engineering of diverse atomic edges and their corresponding morphologies.[76-80] This includes (i) 1D edge nanoribbons by metal particle driven, substrate-facet guided or space-confined growth mechanisms [81-83], (ii) 2D dendrites by chalcogen flux modulation[62, 84-87], and (iii) 3D screw-dislocation spirals on non-Euclidean surfaces[88-92]. Unlike the formation of both zigzag and armchair edges observed in epitaxial graphene or h-BN, CVD-grown TMD monolayers have a unique preference for zigzag edges. The predominant configurations include 1H-zigzag-M(X), 1T-zigzag-M(X), and 1T'-zigzag-M(X), attributable to the inherent instability of armchair edges in monolayers under the prevailing growth conditions. It is reported that the nickel nanoparticle, introduced in the metal particle driven growth of TMDs , would catalyse homoepitaxial growth of a second layer with uniform armchair edges.[93] These principal atomic edge configurations and their corresponding edge morphologies are schematically represented in Figure 1. Despite rapid growth in the field of edge engineering of TMDs by CVD for novel properties, there is a lack of comprehensive and systematic overview on the methods for edge structure manipulation and investigation of the resulting properties in TMDs.

In this review, we aim to provide a comprehensive commentary of the current state of research in the edge engineering of TMDs. The focus of the paper will be on the growth strategies for manipulating edge structures by CVD, the unique properties that result from these manipulations, and the challenges and opportunities for future research in this field. The purpose of this review is to stimulate ongoing research and advancement efforts in utilizing CVD as a scalable method for harnessing the benefits of engineered edge structures in TMDs and exploring their unique properties and applications.

## 2. Atomic Edge Configurations

TMD layers, like graphene and h-BN, exhibit two edge configurations: zigzag and armchair.[59, 94] In figure 2a, the zigzag edge structure consists of atoms arranged in a zigzag pattern with alternating bonds pointing in opposite directions, resulting in a high density of unpaired electrons. In contrast, the armchair edge structure has atoms arranged in a straight line with all bonds pointing in the same direction. Distinct magnetic and electronic properties are predicted for zigzag edge and armchair edge: zigzag edges show the ferromagnetic and metallic behaviour, while armchair edges are nonmagnetic and semiconducting.[59, 95, 96] The presence of a sulfur vacancy in zigzag edges can theoretically lead to the emergence of half-metallicity.[97] However, due to the high-energy requirement, high surface reactivity, slow formation kinetics and high sensitivity to defect or impurities, the armchair edges are often shown to be less stable than the zigzag edges[98]. Thus, zigzag configurations are predominantly observed experimentally, and achieving uniform armchair edges in TMD monolayers is typically challenging.

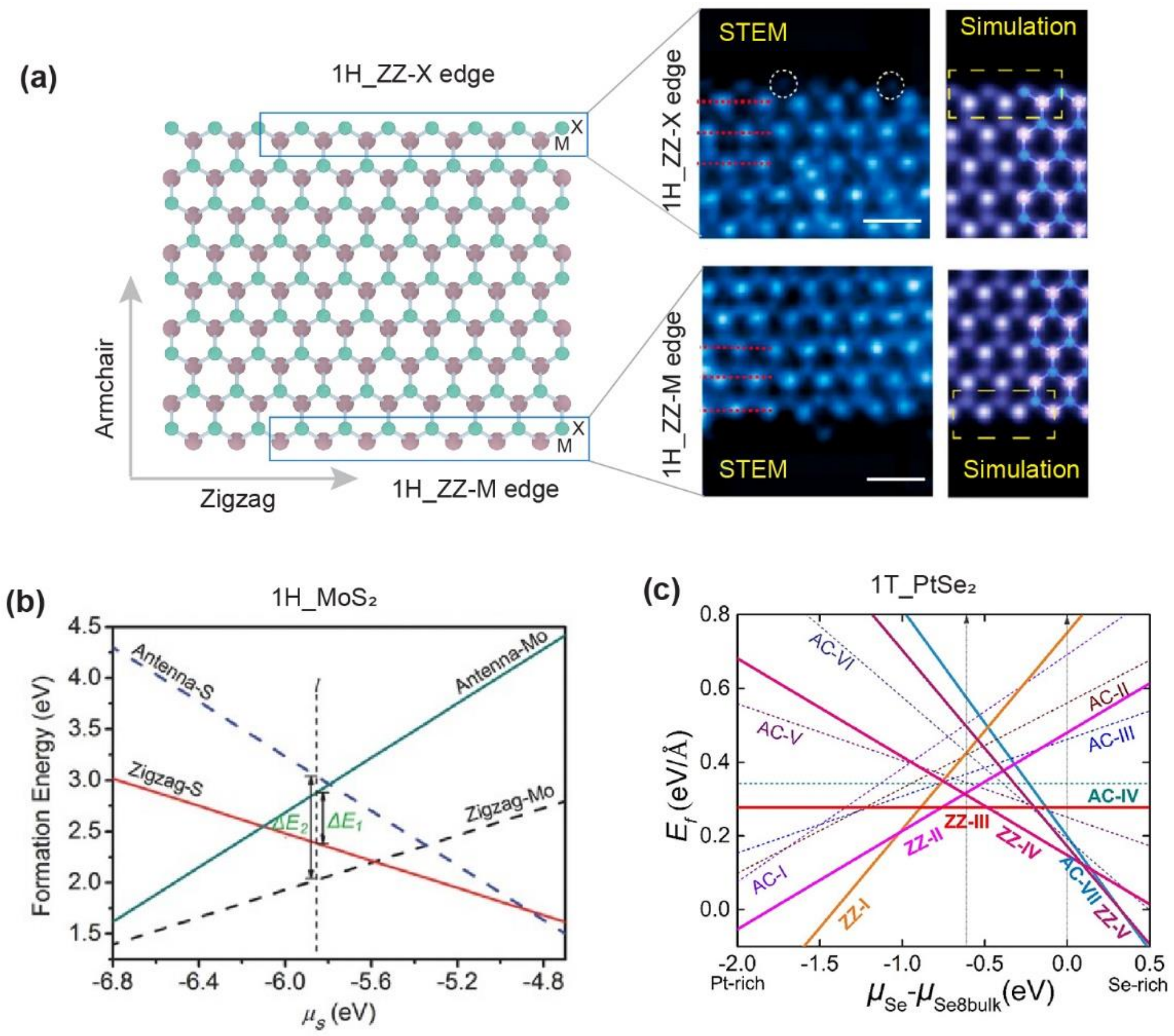

**Figure 2.** (a) Schematic and annular dark-field (ADF) scanning transmission electron microscopy (STEM) images of atomic edge configurations in 1H TMDs. Scale bars: 0.5 nm. (b, c) Variation of the edge formation energy with the chemical potential of sulphur ($\mu_S$) and selenide ($\mu_{Se}$) for the different edge configurations in 1H-$MoS_2$ and 1T-$PtSe_2$, respectively. (a) Reproduced with permission[99]. Copyright 2018, American Chemical Society. (b) Reproduced with permission[62]. Copyright 2018, Wiley-VCH (c) Reproduced with permission[100]. Copyright 2021, Elsevier.

Various fabrication techniques have been employed to create edge-rich TMD layers, like CVD[81, 101], MBE[70, 102], PVD[103], mechanical[104] or chemical exfoliation[105], and wet synthesis[106]. Typically, the as-grown TMD layers are terminated by zigzag edges. By using the step edges of high-miller-index Au facets as templates in the CVD growth, Li et al[81] synthesized arrays of unidirectionally orientated monolayer TMD ribbons with Mo-zigzag edges along the Au<110> steps and free S-zigzag edges on the terraces of the Au surface. Zhao et al[99] employed the MBE technique to grow wafer-scale homogeneous edge-rich nanoporous monolayer $MoS_2$ film with preliminary Mo or S-terminated zigzag edges and reconstructed Mo-terminated edges. Bertoldo *et al.*[103] performed pulsed laser deposition to fabricate $MoS_2$ monolayers with nm-sized grains, revealing atomically sharp zigzag edges. Chemical exfoliation methods were applied to synthesize ultrathin 2H $MoTe_2$ or 1T' $WTe_2$ flakes with rich zigzag edges, promoting the performance in energy storage and electrocatalysis.[105, 107]. Moreover, until now, to the best of our knowledge, the armchair edges of TMD monolayers have been primarily observed in mechanically exfoliated TMD samples. Huang *et al.*[108] conducted tip-enhanced Raman spectroscopy testing on mechanical exfoliated $MoS_2$ flakes, through the double resonance Raman scattering process and electron-phonon interaction in edges, the existing of zigzag and armchair edges in the exfoliated flakes are clearly identified. This was further established by atomic resolution scanning transmission electron microscope (STEM) imaging[109]. Interestingly, introducing nickel particles in the CVD growth of $MoS_2$, Li *et al.*[93] reported the presence of uniform armchair edges formed in a second nanoribbon layer due to the nickel-promoting catalytic metal particle driven growth mechanism, while the first layer still retain the zigzag edges.

Furthermore, several post-treatment techniques have been developed to perform edge engineering in TMD basal layers. Plasma exposure (under $O_2$, Ar or $CHF_3$) or electron beam irradiation can etch the basal planes, creating a porous structure with exposed zigzag edges.[110-112] Through a combination of standard lithography nanofabrication and anisotropic wet etching, precise control over the edges and edge-plane ratio of TMDs was achieved, resulting in nearly atomically sharp and zigzag-terminated.[104] To further control the atom termination in the zigzag edges, Huang *et al.*[113] employed laser irradiation and an improved anisotropic thermal etching process under a determined atmosphere. Their results indicated three possible edge-types may be obtained by controlling the atmosphere: (i) triangular etched hole arrays terminated by the zigzag-W edge in an $Ar/H_2$ atmosphere, (ii) hexagonal etched hole arrays terminated by a mix of zigzag-W and zigzag-S edges in a pure Ar atmosphere, and (iii) triangular etched hole arrays terminated by zigzag-S edges in an Ar/sulfur vapor atmosphere.

Many of the above growth and post-treatment techniques for TMD edge engineering suffer from limitations, such as complexity, potential damage to electronic property under harsh conditions, and lack of precise edge control. As discussed previously, CVD technique shows great potential in this field. The flexibility of varying precursor ratios during growth, allows for tailored atomic edge configurations. The formation of zigzag-terminated edges is strongly influenced by the chemical potential (μ), which is proportional to the ratio of metal to chalcogen (M:X) in the growth environment. For instance, the density functional theory (DFT) model simulations (Figure 2b,c depicts the relationship between edge configurations and the precursor chemical potential, illustrating the variation of edge formation energy with the chemical potentials of sulphur ($\mu_S$) and selenide ($\mu_{Se}$) in 1H $MoS_2$ and 1T $PtSe_2$, respectively.[62, 100] Moreover, in addition to controlling atomic edge configurations, the CVD technique also offers advantages in modulating edge morphologies, such as 1D nanoribbons, 2D dendrites and 3D spirals. By utilizing atomic step engineering of the substate before the CVD growth process, regular 1D nanoribbon edges were obtained with well-aligned orientations.[81, 89] Moreover, through modulating the chalcogen flux during the growth, the edges display the morphology of 2D dendrites.[62] The introduction of the non-Euclidean substrate surface facilitates the fabrication of 3D spiral edges using the screw-dislocation growth mechanism.[88-92] These multi-faceted approaches enable the CVD method a versatile platform for engineering various atomic edges and their corresponding edge morphologies.

## 3. Edge Morphologies

In addition to manipulating atomic sharp edges, the CVD technique serves as a versatile platform for engineering diverse edge morphologies in TMD layers. The morphologies of TMD edges primarily arise from the distinctive shapes of their domains. By employing various growth control mechanisms, the CVD technique enables the formation of TMD edges with different dimensionalities, including 1D ribbons, 2D dendrites, and 3D spirals. This capability expands the range of possible structures and functionalities in TMD materials, enhancing their potential for various applications.

### 3.1. 1D edges from nanoribbon

The formation of TMD 1D edge morphology primarily depends on the controlled growth of 1D TMD nanoribbons. Several growth strategies have been reported to facilitate the growth of TMD nanoribbons. These strategies include the modulated substrate surface-guided growth mechanism, where the surface of the substrate is engineered by atomic steps or chemical modification to guide the growth of nanoribbons.[81, 114] Additionally, the catalytic or noncatalytic metal particle driven growth mechanism and the space-confined growth mechanism have been utilized to promote the formation of 1D TMD nanoribbons.[83, 93]

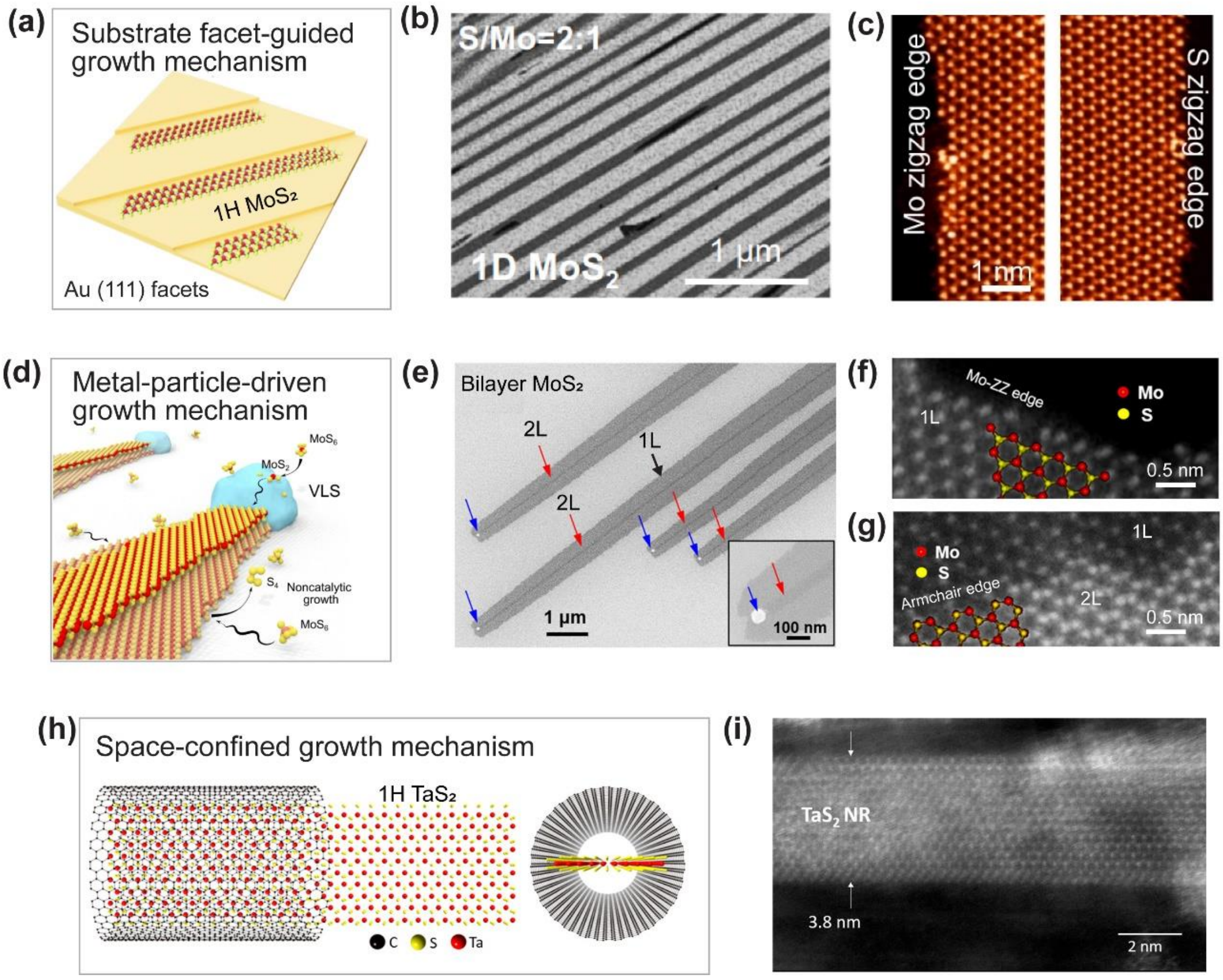

**Figure 3.** 1D edge morphologies. (a) Schematic illustration of the substrate facet-guided growth of well-aligned monolayer $MoS_2$ ribbons along the step edges of vicinal Au(111) facets; (b) SEM images of 2D monolayer $MoS_2$ nanoribbons achieved at an S/Mo ratio of 2:1; (c) Atomic-resolution HAADF-STEM image of $MoS_2$ monolayer ribbon zigzag edges; (d) Schematic of metal particle driven growth of $MoS_2$ ribbons; (e) SEM images of ribbons; (f) Atomic-resolution HAADF-STEM image of zigzag edges located at bottom layer of the bilayer $MoS_2$ ribbons; (g) Atomic-resolution HAADF-STEM image of armchair edges located at top layer of the bilayer $MoS_2$ ribbons; (h) Schematic of space-confined growth of $TaS_2$ nanoribbons templated by carbon nanotubes; (i) High-resolution ADF-STEM image of a $TaS_2$ nanoribbon with ~4nm width. (a-c) Reproduced with permission[81]. Copyright 2022, Springer Nature Publishing AG. (d-g) Reproduced with permission[93]. Copyright 2021, AAAS. (h-i) Reproduced with permission[82]. Copyright 2021, American Chemical Society.

In the epitaxial growth of TMD layers, the step edges of growth substrates can serve as templates for synthesizing TMD nanoribbons with 1D edge morphology. Yang *et al.*[81] developed the novel growth substrate of anisotropic high-miller-index facets Au facets via the melting of Au foils on W templates at 1100 °C, enabling the controllable growth $MoS_2$ nanoribbons. The growth mechanism is schematically illustrated in Figure 3a, and the SEM image in Figure 3b shows the synthesis of large-area uniform, unidirectionally aligned monolayer $MoS_2$ ribbons by using the step edges of vicinal Au(111) single crystals as growth fronts. The widths and lengths of the obtained $MoS_2$ ribbons can be tuned within the ranges of 20–120 nm and 3–30 μm, respectively. Atomic-resolution STEM imaging (Figure 3c) further confirmed the presence of zigzag-Mo and zigzag-S atomic edge configurations on both sides of the monolayer TMD nanoribbon, aligning perfectly with the uniformly orientated step edges of Au. This universal growth strategy is highly promising for controllable growth of wafer-scale, unidirectional 1D TMD edges. Additionally, the chemical treatment of the substrate prior to CVD growth is also shown to be an effective way to control the growth of 1D TMD nanoribbons,

Chowdhury *et al.*[114] conducted CVD growth of $MoS_2$ on Si(001) surfaces that were pre-treated with phosphine, resulting in the formation of high-aspect-ratio nanoribbons of a uniform width. These nanoribbons are reported to exhibit random orientation on the surface with widths ranging from 50 to 430 nm. STEM imaging further revealed the predominant presence of zig-zag edges in the nanoribbons.

Similar to the growth of graphene and BN nanotubes,[115, 116] the metal particle driven growth mechanism has been employed in the growth of TMD nanoribbons. Li *et al.*[93] developed a metal particle driven -related method for nickel particle-enabled width-controlled growth of bilayer $MoS_2$ nanoribbons, as depicted in Figure 3d. They declared that nickel nanoparticles promoted the heterogeneous nucleation of the first layer and catalyses the homoepitaxial tip growth of a second layer, as observed in SEM image in Figure 3e. The width of the bilayer nanoribbons ranged from 8 to 100 nm, mimicking the distribution of Ni particle diameter. Notably, the atomic resolution STEM imaging (Figure 3f,g) revealed that the top-layer nanoribbon exhibited armchair edge configurations, growing in the armchair direction, while the wider bottom layer predominantly exhibited a Mo-terminated zigzag-Mo edge. The top nanoribbon growth followed a typical metal particle driven droplet-catalysed process, where the longitudinal speed was driven by the faster metal particle driven mechanism, while the slower noncatalytic transverse widening of the bottom $MoS_2$ layer occurred with a sufficient supply of feedstock. Further, DFT calculation indicated that the zigzag edges of $MoS_2$ had a stronger affinity for the Ni surface, suggesting the growth direction aligned with the armchair direction. In addition, Li *et al.*[83] reported metal particle driven growth of $MoS_2$ nanoribbons which are several hundred nanometres wide, on NaCl crystal surface. They observed that the alkali metal halide reacted with transition metal oxide precursors to form molten Na-Mo-O droplets that crawled on the substrate surface, mediating the highly anisotropic growth of nanoribbons. STEM imaging revealed the presence of zigzag atomic edges. The inconsistent atomic edge configurations observed in TMD nanoribbons via Ni particle- or NaCl- related metal particle driven growth processes are likely due to the diversity in their catalytic growth abilities.

Additionally, the space-confined growth strategy has been reported for TMD nanoribbon growth. Carbon nanotube (CNT) reaction vessels serve as templates for the selective grow of 1D TMD nanoribbons with precise width control[82, 117, 118], as depicted in Figure 3h. The growth method involves depositing stoichiometric elemental precursors alongside partially opened CNTs, the mixture is then heated at temperatures at 600-900 °C over several days, followed by a significant cooling time, resulting in TMD nanoribbons. The STEM image in Figure 3i shows a rough bare-zigzag atomic edge configuration with a slight offset from the lattice direction. The geometrical confinement within the interior nanotube induced compressive strain on the encapsulating CNTs[82], leading to a strong interaction between the edge of the TMD nanoribbon and CNT interior. This interaction facilitates charge transfer from the CNT to nanoribbon through the zigzag edges, favouring formation of the 1H phase due to the ability of its available valence bands crossing the Fermi energy to host excess electrons. However, this nanoribbon growth method is subject to limitations such as synthesis complexity and poor width control.

Moreover, The TMD nanoribbons can also be produced through direct chemical[119, 120] or laser-induced[121] unzipping of TMD nanotubes. For instance, multilayer $WS_2$ nanoribbons can be synthesized by intercalating Li, K, or Na into multiwalled $WS_2$ nanotubes, followed by a reaction between the intercalated metal and long-chain thiols[119, 120]. This method enables large-scale fabrication of TMD nanoribbons with controllable widths and high yield. However, the rigorous reaction process makes it challenging to obtain monolayer nanoribbons with precisely controlled sharp atomic edges.Hence, the step edges-guided growth strategy shows significant potential for the controlled synthesis of large-scale, relatively uniform, and unidirectionally oriented 1D TMD zigzag edges. Moreover, the metal particle driven growth strategy enables the catalytic growth of 1D atomic edges, including armchair configurations. Additionally, the CNT-space-confined growth method offers a means to fabricate 1D nanoribbon edges with precise control over their width and strong interfacial interactions with CNTs.

It should be emphasized that in edge-rich 2D nanoribbons, the edge effect can be significant and become dominant. Interestingly, several edge effects, previously anticipated only in theory, have now been experimentally realized. Examples of these are the observation of spin splitting of dopant edge states[38], long-range nontopological edge currents[122], and fractional edge excitations[123]. In addition to the width of the nanoribbon, the fine control of atomic edge types and the creation of specific superlattices with epitaxial substrates have significant influence on their electronic properties.[40, 124, 125] Considering these insights, the method of substrate-facet guided nanoribbon growth of TMD offers potential benefits. This technique affords robust control over width, stacking order, and step-dominant alignment of nanoribbons, making it a promising strategy for the design of distinct TMD nanoribbons with unique properties.

### 3.2. 2D edges from fractals, dendrites and compact domains

The 2D edges of TMD layers primarily reside within the grown TMD domains in a fractal-dendritic-compact morphological range. The density and morphology of these 2D edges are strongly influenced by the fractal nature of the domains formed during the CVD process. Fractals follow a growth mechanism known as diffusion-limited aggregation, while dendrites exhibit preferential growth along symmetry-governed directions. On the other hand, the formation of straight-edged compacts is governed by thermodynamic mechanisms. The shape evolution of these structures is determined by the interplay between adatom attachment to the domain edges and the diffusion of adatoms along the domain boundaries.

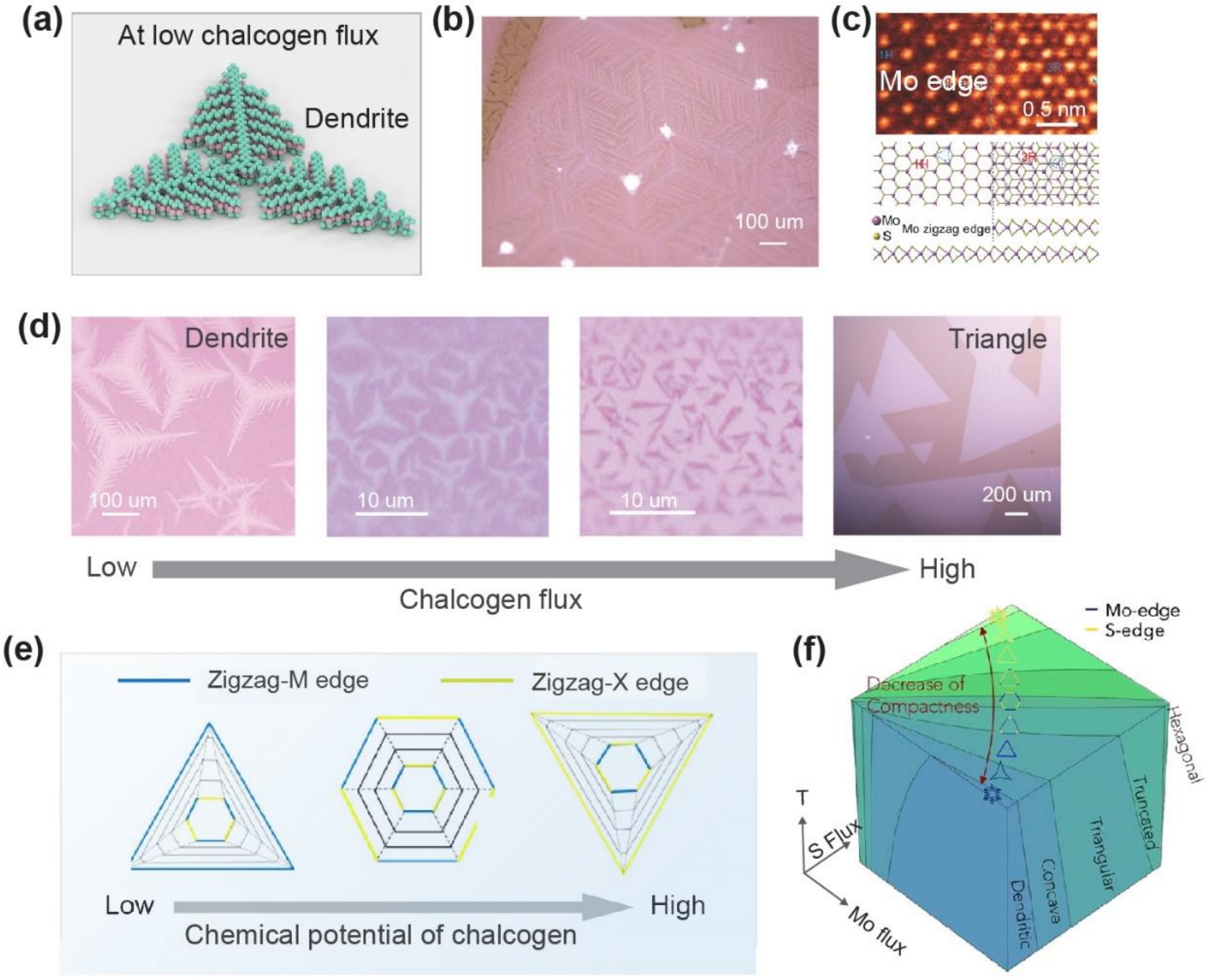

**Figure 4.** 2D edge morphologies. (a) Schematic of dendritic TMD flake; (b) Optical image of highly organized dendritic $MoS_2$; (c) STEM-ADF image of a typical $MoS_2$ region showing a second layer with a ZZ-Mo edge; (d) The chalcogen flux dependent domain shape evolution from dendrite to triangle; (e) Chemical potential of chalcogen dependent atomic edge configuration in compact TMD layers; (f) Schematic illustration of a complete edge morphology

diagram with Mo flux, S flux and growth temperature. (b-d) Reproduced with permission[62]. Copyright 2018, Wiley-VCH; (e) Adapted with permission[87]. Copyright 2014, American Chemical Society. (f) Reproduced with permission[126]. Copyright 2020, Elsevier.

Several control strategies have been developed to grow the high-fractality domains with rich exposed edges by promoting diffusion-limited growth and limiting thermodynamic growth.[127] One straightforward approach involves suppressing thermodynamically favoured edge diffusion by conducting CVD growth at lower temperatures. Zhang *et al.*[128] and Chowdhury *et al.*[129] demonstrated that the shape of the $MoS_2$ or $MoSe_2$ domain can transition from compact to fractal and dendritic structures as the growth temperature is reduced. Moreover, the precursor flux ratio also plays a key factor in controlling the domain shapes in the growth. Higher metal flux or lower chalcogen flux increases the number of preferred growth directions and promotes the anisotropy of edge diffusion. The adatoms, in addition to attaching to the edges, preferentially diffuse along the lowest-energy paths on the substrate, resulting in branching along the directions of the domain expansion and the formation of dendritic TMD domains, as illustrated in Figure 4a. Chen *et al.*[62] reported the homoepitaxial growth of large-scale dendritic $MoS_2$ patterns on a $MoS_2$ monolayer via chemical vapor deposition. At low sulfur flux, regular trigonal symmetric dendrites patterns are clearly observed in a large scale, as shown in Figure 4b. The atomic resolution STEM imaging in Figure 4c revealed that the dendrites edges are predominantly terminated by zigzag-Mo edge structure. With the chalcogen flux increases, the dendrites structures transform into compact triangles, as shown in Figure 4d. The pattern formation can be attributed to the anisotropic growth rates of the zigzag-S and zigzag-Mo edges under low sulfur chemical potential.

In the growth process, as the domains approach fully thermodynamic growth and transform into compact shapes, the increasing chalcogen chemical potential becomes crucial in determining both the atomic edge configuration and the formation of the compact structures. As schematically shown in Figure 4e, initially, at low chalcogen chemical potential in a Mo-rich environment, zigzag-S edges grow faster than the zigzag-Mo edges, and would disappear, leaving a triangle with three zigzag-Mo edges. When the sulfur chemical potential reaches to the stoichiometric ratio of $MoS_2$, the zigzag-Mo and zigzag-S edges show similar stability and growing rate, resulting in a hexagonal shape with three zigzag-Mo edges and three zigzag-S edges. With further increase in the sulfur chemical potential, the zigzag-Mo edges become unstable and vanish during growth, ultimately leading to a triangular structure of three sides of zigzag-S edges. Furthermore, Zhang *et al.* [126] developed the morphology diagram that depicts the evolution of 2D TMD domain shapes based on the growth parameters of Mo or S flux ratio and growth temperatures, as illustrated in Figure 4f. This morphology diagrams allow for a comprehensive understanding of crystal shape evolution over time and provide insights into the relationship between growth conditions and resulting morphology.

### 3.3. <u>3D edges from spirals and nanotubes</u>

The 3D edges of TMD layers primarily originate from their corresponding 3D spirals and 1D nanotubes. The controllable growth of 3D spiral domains is achieved via screw-dislocation-driven (SDD) growth mechanism, which has been successfully employed in synthesizing various layered spiral materials using different fabrication techniques.[90, 92, 130] For example, $MoS_2$[89, 131], $WS_2$[88], $WSe_2$,[132] $MoTe_2$[133], graphene[134], BN[135], SnS[136] have been synthesized via CVD. Additionally, aqueous solution synthesis has been utilized to fabricate conjugated block copolymers[137], CoAl layered double hydroxide[138] and $Ni(OH)_2$[139]. Furthermore, various strategies have been developed for the fabrication of 1D TMD nanotubes. These include hydrothermal synthesis, epitaxial growth of 1D TMD nanotubes on templating carbon or BN nanotubes[140-143], and the capillary-force-driven rolling up process that leads to the delamination of TMD lateral domains and the formation of 1D TMD nanotubes[144]. In this section, our discussion mainly focuses on epitaxial growth of 3D edges by CVD.

In layered materials, the presence of a screw dislocation adds an intriguing dimension to their structure. A screw dislocation represents a singularity in space that connects different layers of the 2D crystal structure in a continuous spiral. This arrangement introduces additional complexity in the stacking of layers, going beyond the typical few-layer level. Spirals, in particular, exhibit a distinct lattice structure where a single overgrown monolayer stacks on top of itself in a circular pattern, gradually decreasing in width to form a pyramidal morphology. During growth, the formation of a domain occurs through diffusion-limited growth mechanism, when two edges within the same domain meet with a lattice mismatch, one of the grain boundaries lifts over the other, resulting in a line of unsaturated chalcogen dangling bonds on the lifted edge, forming dimers. At the end of this line, a screw dislocation emerges, facilitating further spiral growth, as depicted in Figure 5a. Interestingly, screw dislocations can be intentionally induced experimentally by increasing the likelihood of misaligned meeting between two exposed edges. As the two-dimensional growth rate increases, these accidental overgrowths occur with greater regularity. Consequently, the generation of a screw dislocation relies on three essential growth factors: (i) differential growth rates between the incident edges within the domain to induce edge lifting, (ii) a chalcogen-rich environment for the formation of linearly aligned metal vacancies, and (iii) a high growth rate that favors the rapid growth of certain edges over slower and more stable ones. The growth parameters, such as precursor supersaturation, significantly influence the spiral growth. As growth progresses around the screw dislocation, the exposed edges follow the edge corners of the underlying layers. The in-plane size of the domain, also known as the terrace width, is primarily determined by reactant concentrations, which are influenced by the lateral step velocity and the out-of-plane growth rate. The lateral step velocity corresponds to the typical horizontal growth of monolayers, while the out-of-plane growth rate represents the speed at which subsequent layers form as a full rotation around the screw dislocation takes place.

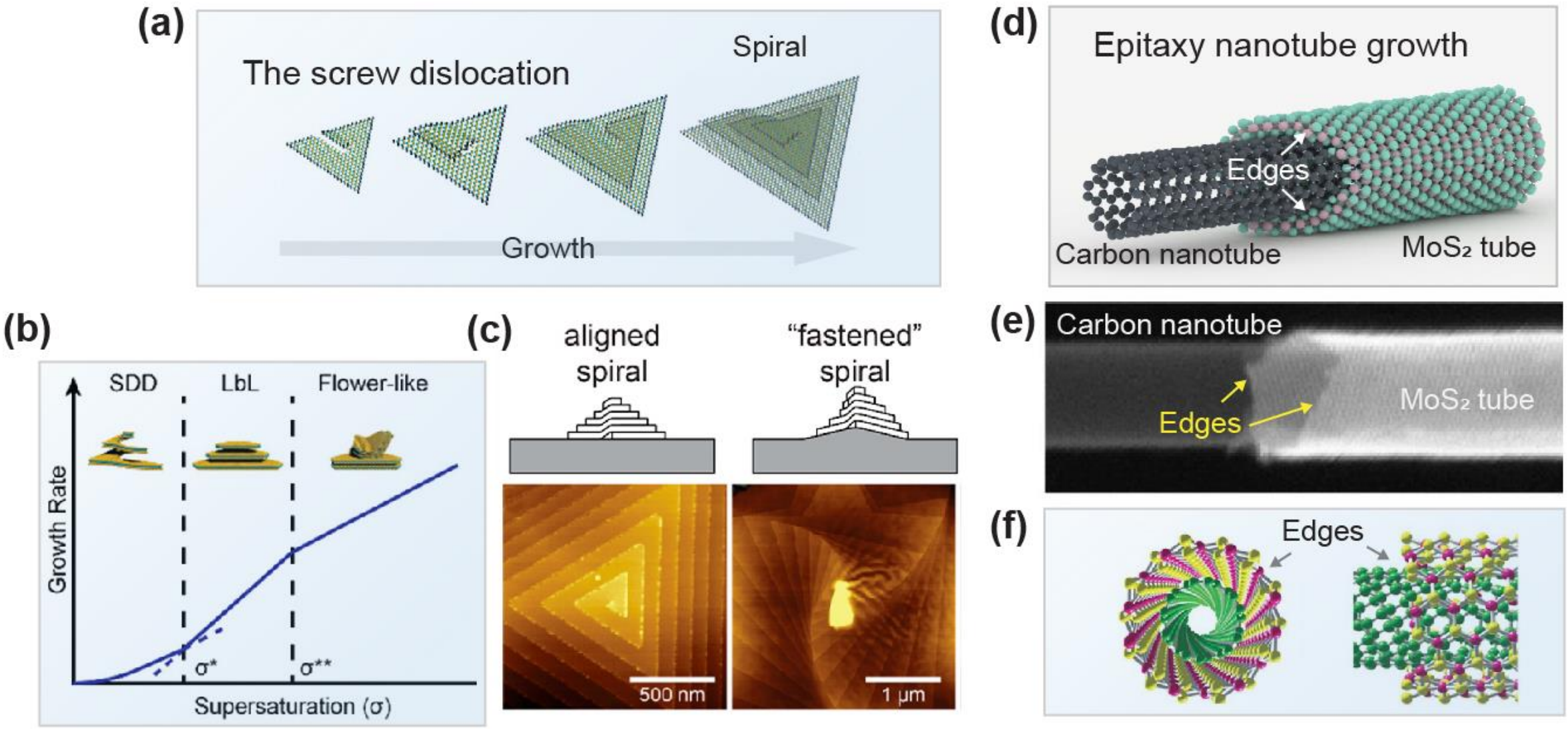

**Figure 5.** 3D edge morphologies. (a) Schematic of TMD spiral growth by the screw dislocation driven mechanism. (b) Schematic illustrations of supersaturation dependent the growth rate of different dominant domain growth modes of screw dislocation driven (SDD), layer-by-layer (LbL) and flower-like mechanisms; (c) AFM images of aligned and supertwisted spiral $WS_2$ growths on $SiO_2$/Si substrates. (d) Schematic of an epitaxial single-walled TMD grown on a carbon nanotube; (e, f) Atomic resolution STEM imaging and schematic of a grown single-walled $MoS_2$ nanotube edges; (a) Adapted with permission[145]. Copyright 2018, American Chemical Society. (b) Adapted with permission[92]. Copyright 2022, American Chemical Society. (c) Reproduced with permission[89]. Copyright 2020, AAAS. (e,f) Reproduced with permission[142]. Copyright 2020, AAAS.

Supersaturation, defined as the logarithm of the precursor relative to the equilibrium concentration, plays a crucial role in crystal growth. Increasing supersaturation levels lead to the dominance of different growth modes, such as SDD growth, layer-by-layer (LbL) growth, and flower-like dendritic growth (Figure 4b). A supersaturation above the critical supersaturation ($\sigma^*$) is necessary to overcome the nucleation energy barrier for LbL growth to generate a new layer. However, the presence of a screw dislocation can promote crystal growth under lower supersaturation conditions, this is because the screw dislocation, upon intersecting with the crystal surfaces, inherently possesses step edges without the need to overcome the energy barrier associated with nucleating new crystal steps. Therefore, crystal growth can still occur below the critical supersaturation level. Fan *et al.*[145] conducted a study on the controllable growth of multi $WS_2$ spirals. By adjusting the CVD growth conditions, such as the growth temperature, the supersaturation of the crystal growth can be modulated. As a result, different types of spirals can be grown, including multilayer normal vdW layers through LbL growth mechanism, dislocated triangular spirals with an increasing number of dislocations per domain, and triangular-hexagonal mixed spirals. The transition from triangular spirals to hexagonal spirals occurs through the introduction of new screw dislocations, leading to the formation of complex multi-dislocation spirals. Shearer *et al.*[132] also demonstrated that screw dislocations have the ability to modify the shapes of $WSe_2$ spirals. This modification is governed by the rotation and number of elementary triangular dislocation spirals. The symmetry of the dislocation spirals determines three types of spiral morphologies of triangular, hexagonal, and mixed.

Additionally, a growth strategy involving non-Euclidean surfaces has been employed to achieve the formation of screw-dislocated spiral superstructures with continuous interlayer twists.[89] In traditional crystal lattices following Euclidean geometry, the angular period is typically 360° (Figure 5c, aligned spiral). However, when a layered crystal grows on a non-Euclidean (curved) surface, the angular period can be mismatched, resulting in a larger or smaller value than 360°. This mismatch between the non-Euclidean surface and the Euclidean lattice leads to an additional lattice twist for each successive period, creating a consistent and continuous twist angle (Figure 5c, fastened or supertwisted spiral). Experimental demonstrations of this non-Euclidean twist phenomenon have been carried out in the vapor-phase growth of $WS_2$ and $WSe_2$ on substrates modified with appropriate geometric features. By using deformed $SiO_2$ nanospheres, drop-casted $WO_3$ nanoparticles, or spontaneously grown $WO_x$ nanoparticles as protrusions, it was possible to influence the centre of triangular and hexagonal TMD spirals with non-Euclidean geometry, resulting in the continuous generation of supertwisted layers. This approach highlights the intriguing effects of non-Euclidean surfaces on the growth and structure of spiral superstructures in TMDs.

By using carbon or h-BN nanotubes as template, the single-walled, single-crystal TMD nanotube can be epitaxially grown on the outer of carbon or h-BN nanotubes, following an open-end growth mechanism.[146-150] The nucleation and extension of epitaxial TMD nanotube occur exclusively at the open edge of the carbon nanotube, as schematic illustration in Figure 5d. Typically, two circular or elliptical closed edges are formed at the ends of the TMD nanotube, with the diameter of these edges strongly dependent on the diameter of underlying carbon nanotube. For instance, the $MoS_2$ nanotubes with diameters ranging from 3.9 to 6.8 nm are observed to grow coaxially on carbon nanotubes. Xiang et al[142]. Reported that for 5-nm-diameter monolayer $MoS_2$ nanotube coaxially grown on carbon nanotube, atomic resolution STEM imaging (shown in Figure 5e) clearly showed the elliptical closed edges. Similarly, Furusawa et al[151]. used h-BN nanotube as template and achieved epitaxial growth of 4.4-nm-diameter single-wall $MoS_2$ nanotubes, where the zigzag edge was identified through atomic resolution STEM image. These nm-sized closed edges may exhibit unique quantum confinement effect on advanced electronic, spintronic, or optoelectronic properties. It is worth noting that this controllable growth method, which enabled access to these few-nm-diameter-sized circular edges, is distinct from other methods that involve rolling up lateral grown domains to form 1D nanotubes.

Table 1 Overview of edge engineering in atomic TMDs by CVD and their corresponding edge specifications

| Edge morphology | | Material platform | Growth strategy | Growth parameters | | Edge control | | | | Ref. |
|---|---|---|---|---|---|---|---|---|---|---|
| | | | | Substrate | Precursors | Primary edge type | Grain size | Edge perimeter/area | Edge atom density/area | |
| 1D | nanoribbon | 1H-$MoS_2$ | Metal particle driven | NaCl, $SiO_2$/Si, sapphire, $MoS_2$ | $Na_2MoO_4$, $MoO_3$, S | 1H-ZZ-Mo, 1H-ZZ-S | W: 50-1100 nm; L: 5-320 µm | 1.82-40.20 $\mu m/\mu m^2$ | 1.82 $x10^4$-4.0 $x10^5$ $\mu m^{-2}$ | 83 |
| | | 1H-$MoS_2$ | Metal particle driven | $SiO_2$/Si, $MoS_2$ | $MoO_2$ + Ni + NaBr, S | 1H-ZZ-Mo, 1H-ZZ-S, 1H-armchair | W: 8-100 nm; L: 2-70 µm | 20.03-251 $\mu m/\mu m^2$ | 2.0 $x10^5$-2.5 $x10^6$ $\mu m^{-2}$ | 93 |
| | | 1H-$MoS_2$ | Step edges as templates | Aµ high-index facet | $MoO_3$, S | 1H-ZZ-Mo, 1H-ZZ-S | W: 20-120 nm; L: 3-30 µm | 0.06-100.67 $\mu m/\mu m^2$ | $6x10^2$-$10^6$ $\mu m^{-2}$ | 81 |
| | | 1H-$MoS_2$ | Surface-modification | $P_x$-Si(001) | $MoO_3$, S | 1H-ZZ-Mo, 1H-ZZ-S | W: 50-430 nm; L: 150 µm | 4.4-399 $\mu m/\mu m^2$ | 4.4 $x10^4$-$4.0x10^6$ $\mu m^{-2}$ | 114 |
| | | 1H-$TaS_2$ | Space-confined | Carbon nanotube | Ta, S, $I_2$ | 1H-ZZ-Ta, 1H-ZZ-S | W: 2.5-5 nm; L: 2.5 µm | $4.2x10^2$-$10^3$ $\mu m/\mu m^2$ | $4.2x10^6$-$10^7$ $\mu m^{-2}$ | 82 |
| | | 1T' $WTe_2$ | Promoter-assisting | $SiO_2$/Si | Te and mixed compounds of $WO_3$, $WCl_4$, and Te | 1T' ZZ-W, 1T' ZZ-Te | W: 20 µm; L: 320 µm | 0.11 $\mu m/\mu m^2$ | $1.1x10^3$ $\mu m^{-2}$ | 152 |
| 2D | Dendrite /fractal | 1H $MoS_2$ | Diffusion-limited | Melted glass, sapphire, $SrTiO_3$ | Mo foil, S | 1H ZZ-Mo | 250-350 µm | 0.12-0.14 $\mu m/\mu m^2$ | $1.2x10^3$-$1.4x10^3$ $\mu m^{-2}$ | 62 |
| | | 1T' $MoTe_2$ | T-dependent anisotropic growth | $SiO_2$/Si | $MoO_3$, Te, NaCl | 1T' ZZ-Mo | 11-20 µm | 0.67-0.75 $\mu m/\mu m^2$ | $6.7x10^3$-$7.5x10^3$ $\mu m^{-2}$ | 153 |
| | Compact | 1H $MoS_2$ | Flux control | $SiO_2$/Si, sapphire, graphene, mica, quartz | $MoO_3$, $WO_3$, S, Se, NaCl | 1H-ZZ-Mo, 1H-ZZ-S | 2-1100 µm | 0.0063-3.45 $\mu m/\mu m^2$ | 63 -$3.45x10^4$ $\mu m^{-2}$ | 154 |
| | | 1T $MoTe_2$ | Growth quench temperature | $SiO_2$/Si | $MoO_3$, Te | 1T' ZZ-Mo, 1T' ZZ-Te | 10 µm | 0.4 $um/um^2$ | $4x10^3$ $um^{-2}$ | 84 |
| | | 1T' $MoS_2$ | introduction of potassium | $SiO_2$/Si | $K_2MoS_4$, K, S | 1T' ZZ-S | 2-10 µm | 0.69-3.45 $um/um^2$ | $6.9x10^3$-$3.45x10^4$ $um^{-2}$ | 155 |
| 3D | Spiral | 2H $WS_2$ 2H $WSe_2$ | Screw dislocation driven | $WO_3$ or $SiO_2$ particle coating $SiO_2$/Si | $WSe_2$, $WS_2$, $CaSO_4{\cdot}2H_2O$ | 1H ZZ-W, 1H ZZ-S, 1H ZZ-Se | 3-30 µm | 9.2 $um/um^2$ | 9.2 $x10^4$ $um^{-2}$ | 88, 89 |
| | | 1T' $MoTe_2$ | Screw dislocation driven | $SiO_2$/Si | $(NH_4)_2MoO_4$, NaCl, Te | 1T' ZZ-Mo, 1T' ZZ-Te | ~9 µm | 38.3 $\mu m/\mu m^2$ | 3.8 $x10^5$ $\mu m^{-2}$ | 133 |
| | Nano tube | 1H $MoS_2$ | Single-wall carbon nanotube as a template | Single-wall carbon nanotube | $MoO_3$, S, | 1H ZZ-Mo | Diameter: 6 nm | 80 $\mu m/\mu m^2$ | $8x10^5$ $\mu m^{-2}$ | 142, 151 |

## 3.5 Summary of growth strategies for edge engineering

The manipulation of edges in graphene has garnered significant attention, and similar potential exists in TMD materials. TMDs exhibit great promise due to their versatile chemical compositions, phase structures (2H, 1T, and 1T'), as well as the presence of various edge configurations, variable edge densities, and edge morphologies in 1D, 2D, and 3D forms. Among the various growth techniques, the CVD technique stands out as an effective method for controlling atomic edge types, density, and morphologies in TMDs. Table 1 provides an overview of growth strategies towards TMD edge engineering using CVD.

For the growth of 1D edges in TMD layers, the metal particle driven strategy has proven effective in controlling the width of nanoribbons by utilizing droplet size as a confining factor. This approach leads to a high density of edge atoms per unit area, typically around 2.0 x $10^5$ $\mu m^{-2}$. It should be noted that while the irregular diffusion behaviour of droplets during growth may result in the absence of unidirectional orientation, the overall distribution of 1D edges remains uniform. The scalability and reproducibility limitations associated with droplet-based approaches highlight the need for alternative strategies to control 1D nanoribbon edges. Surface engineering techniques, such as substrate step engineering and chemical surface modification, have emerged as promising methods for synthesizing large-scale, high-density 1D nanoribbon edges. By precisely controlling the growth time and width, it is possible to achieve a controllable edge density of up to 6.0 x $10^6$ $\mu m^{-2}$ with excellent uniformity. Notably, the step edge-guided method enhances the unidirectional orientation of nanoribbons, providing additional control over their growth. The carbon nanotube space-confined growth method exhibits the highest edge atom density of 4.2 x $10^7$ $\mu m^{-2}$. However, its scalability is hindered by the low production yield resulting from the high diffusion barrier of precursors within the narrow-diameter carbon nanotubes.

In terms of 2D edge control, dendritic edges formed through various methods exhibit similar atom densities of several thousand atoms $\mu m^{-2}$ in regular trigonal symmetric dendrite patterns. Strategies based on diffusion control show potential for scaling up the production process. The edge density of straight edges in compact triangles or hexagons is highly dependent on the size of the domains. As the domain size increases from 2 μm to 1100 μm, the edge density decreases from 3.4 x $10^4$ $\mu m^{-2}$ to 63 $\mu m^{-2}$.

For 3D edges, the screw dislocation-driven spiral growth at low supersaturation conditions enables the creation of unique, continuous helical edges with ultrahigh atom edge densities reaching several numbers of $10^5$ $\mu m^{-2}$. The epitaxial TMD nanotubes grown on several-nm-diameter carbon nanotubes also exhibit ultrahigh atomic density of around 8x$10^5$ $\mu m^{-2}$ in closed circular-like edges. However, due to presence of high strains on nm-diameter nanotube surface, the production yield is around 10%, limiting its scale-up fabrication.

Although significant progress has been made in current research, particularly in the controllable growth of various TMD edge morphologies and atomic edge types. However, there still exists a considerable gap in exploring atomic precision engineering of TMD edges and investigating their interlayer interaction with the underlying substrate.

The versatile modulation of edges in epitaxial TMD growth using CVD techniques holds great promise for novel applications in the fields of electronics, optoelectronics, catalysis, and quantum technologies and we shall explore some key ones in the next section.

## 4. Edge Applications

The CVD-based edge engineering and manipulation of various TMDs have unveiled a notable level of complexity in the control of their nuanced structural properties. These intricacies contribute to a wide array of distinctive features and properties that stem from the presence of edge states and the remarkably high edge density inherent to these novel structures. The exploration of these complexities presents exciting prospects for advancing our understanding and harnessing the potential of TMDs in diverse fields, including electronics, optoelectronics, catalysis, and quantum application.

### 4.1. Electrical

The presence of edge structures in TMDs introduces broken spatial symmetry and quantum confinement effects, resulting in distinct electronic properties. Specifically, the band gaps of edge states located at various morphologies are significantly smaller than those of the in-plane counterparts.[156] In 1H-armchair nanoribbons, which possess a direct band gap, the electronic properties exhibit weak dependence on the ribbon width, making them suitable as semiconductors. In contrast, ZZ-terminated nanoribbons exhibit metallic behavior, with asymmetric majority and minority spin bands. These nanoribbons display metallic characteristics with multiple energy levels crossing the Fermi level, effectively closing the band gap.[156] The presence of bare ZZ edges enhances the metallic feature, with a higher number of energy bands crossing the Fermi level (in Figure 6a).[157] Different configurations of ZZ edges exhibit variations in metallic properties, with bare ZZ-M and ZZ-X edges displaying the most pronounced metallic behavior. Furthermore, the edges of 1H-$MoS_2$ undergo a charge reconstruction process, leading to the formation of electron and hole pockets of free carriers. These free carriers form 1D metallic wires extending along the nanoribbon edges, with electrons primarily localized at the zigzag-Mo and holes at the zigzag-S (in Figure 6b).[157]

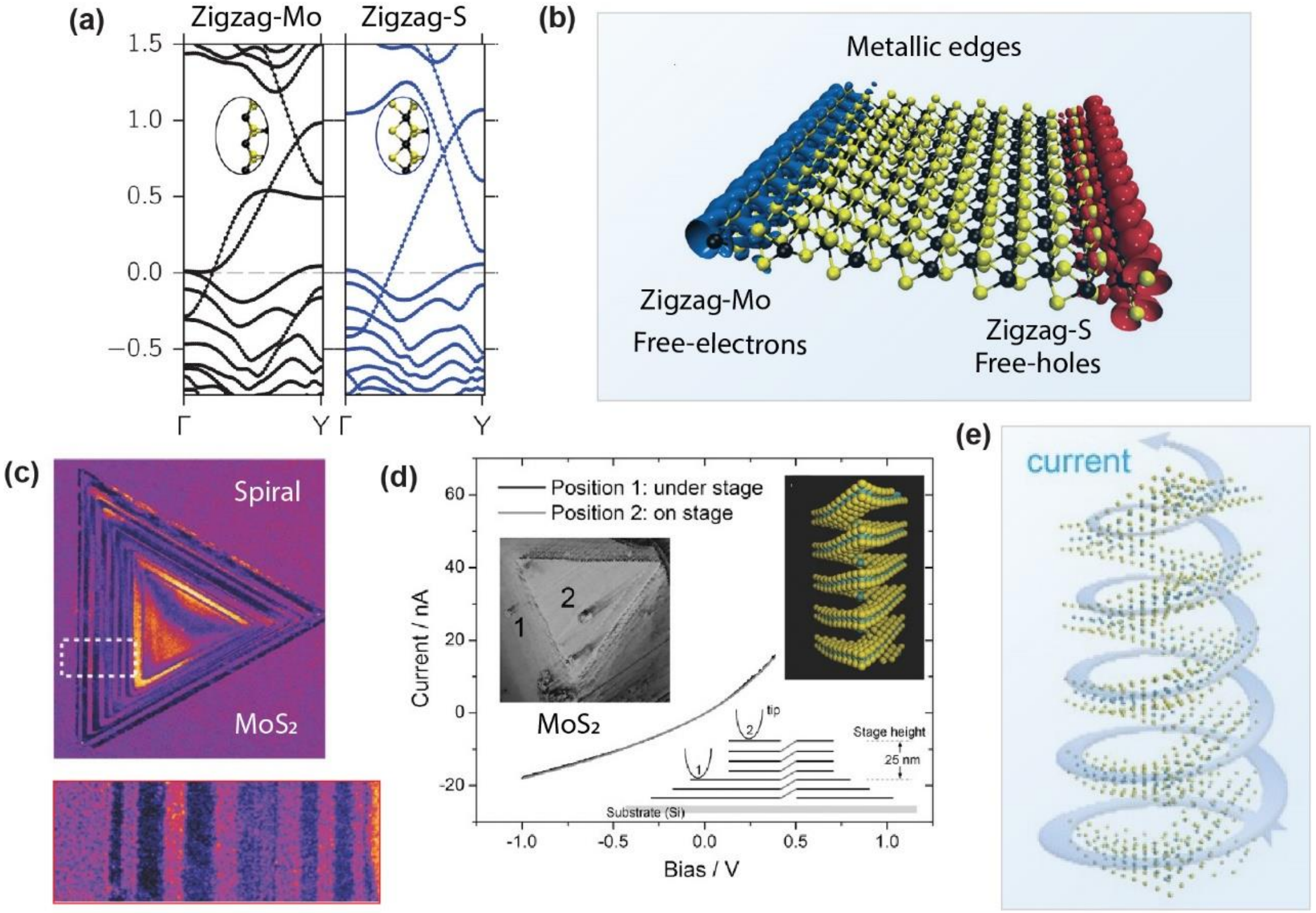

**Figure 6.** (a) Energy band structures for $MoS_2$ nanoribbons with ZZ-Mo and ZZ-S. (b) Real-space distribution of the charge density of free electrons (blue) and holes (red). (c) TEM characterization of $MoS_2$ spiral (d) Two-terminal IV curve for $MoS_2$ spiral measured by c-AFM at two positions. (a, b) Reproduced with permission[157]. Copyright 2015, American Chemical Society. (c-e) Reproduced with permission[131]. Copyright 2016, Wiley-VCH.

The enhanced metallic property in edge structure has also been reported in 1T and 1T’ phases of TMD layers. Liu *et al.*[158] theoretically identified two parallel metallic and magnetic edges located in a 1T-$PtS_2$ nanoribbon. The zigzag-Pt edge state was found to be relatively stable, exhibiting metallic and magnetic properties, mainly contributed by the 5d orbits of Pt atoms at edges. Similar metallic edges have been found in $PtSe_2$ as well.[159] Edges in the 1T' phase of TMDs also exhibit unique electronic properties. Xu *et al.*[160] studied 1T' $MoS_2$ and observed metallic behaviour with a gapless "V-shaped" dI/dV spectrum centred at the Fermi level in 1T'-$MoS_2$ islands at 77.8 K. However, upon cooling the sample to 4.5 K, a gap opening at Fermi level was observed. The dI/dV spectra taken at the center of 1T' islands exhibited a small gap at Fermi level, contrasting the sharp V shape observed at 77.8 K. Interestingly, the metallic behaviour of the edge was preserved in 1T'-$MoS_2$ islands, which can be attributed to the one-dimensional (1D) nature of the edge states. The observed gap opening in 1T'-$MoS_2$ islands, as well as in monolayer 1T'-$WTe_2$ islands[161] and few-layer 1T'-$MoTe_2$ films[162], can be explained by the breaking of band degeneracy due to the effects of spin-orbit coupling in the distorted 1T' phase. These findings highlight the intriguing electronic properties exhibited by edge states in different phases of TMDs. The presence of metallic and magnetic edges in specific TMD materials holds great potential for the design of innovative electronic and spintronic devices.

In metallic 1H TMD layers, such as 1H-$NbSe_2$, distinct 1D electronic states are also present along the edges. Zhang *et al.*[163] investigated the electronic properties of monolayer $NbSe_2$ and observed the emergence of obvious 1D electronic states along the zigzag edges. The charge densities and electronic band structures at the zigzag edges of monolayer $NbSe_2$ revealed significant electronic densities for zigzag-Se edges at the Fermi level. The zigzag edges in monolayer $NbSe_2$ showcase intriguing termination-dependent one-dimensional edge states, accompanied by an energy-dependent charge-density-wave modulation near the edges transitioning from a 3 × 3 charge density wave order to a nearly stripe phase.

The edge morphologies, as well as the atomic edge configurations, play significant roles in influencing electronic properties. For instance, as shown in Figure 6c, 3D spiral edge structures of 2H $MoS_2$[131], which are developed based on the semiconducting nature of 1H domains, exhibit unique features, such as vertical conductance, distinguishing them from regularly stacked multilayer TMDs. In contrast to the high resistivity expected between regularly stacked multilayers—where the van der Waals-bonded layers create conditions unsuitable for interlayer electron transfer—the Frank-Read core of a screw dislocation spiral structure enables a pseudo-interlayer electron pathway. This occurs while all layers of the structure remain interconnected within a warped monolayer plane. Interestingly, almost negligible resistance is observed between positions on an $MoS_2$ spiral, even with significant height separation (25nm, equivalent to approximately four atomic layers). Moreover, the electrical conductance is exceptionally high in the thicker central regions of the spiral, while lower conductance values are found on the thinner outer edges (in Figure 6d,e). The vertical conductance feature offered by these spirals unlocks new possibilities for TMDs in future electronic applications. It presents a valuable asset that can be harnessed to enhance the performance of electronic devices utilizing these versatile materials.

### 4.2. <u>Optical Properties</u>

The edges of TMD monolayers, exhibiting unique electronic structural changes, give rise to enhanced nonlinear optical properties[164-167],, particularly in second-harmonic generation (SHG). This enhancement is attributed to a two-photon resonance effect, which occurs due to subband transitions between the valence bands and the localized edge states. The broken translational invariance at the edges of TMD monolayers is responsible for the emergence of these localized mid-gap electronic states, leading to the significant enhancement of SHG. Multiphoton laser scanning microscopy is utilized to study atomic edges with localized electronic states, which result in greatly enhanced optical SHG with double resonance peaks and an edge-atom dependent difference in the resonance energy. Lin *et al.*[168] observed the increased SHG enhancement at a pump wavelength of 1270 nm for $MoS_2$ zigzag-S edges, which corresponds to the resonant wavelength of the edge states (Figure 7a). Similarly, Yin *et al.*[32] investigated the SHG properties of zigzag-Mo edges and found pronounced enhancement at the edges when pumped at 1300 nm (Figure 7b). The difference in resonance wavelength between zigzag-S and zigzag-Mo edges can be attributed to the energy variations between the lower energy of the zigzag-Mo edge states within the band gap and the higher energy of the zigzag-S edge states. In quantum mechanics, when the additional energy levels of edge states closely coincide with virtual transition levels, the second-order optical susceptibilities are significantly amplified. Therefore, the resonant behavior observed in the SHG response of TMD monolayer edges is a consequence of this energy matching and results in the enhanced nonlinear optical properties.

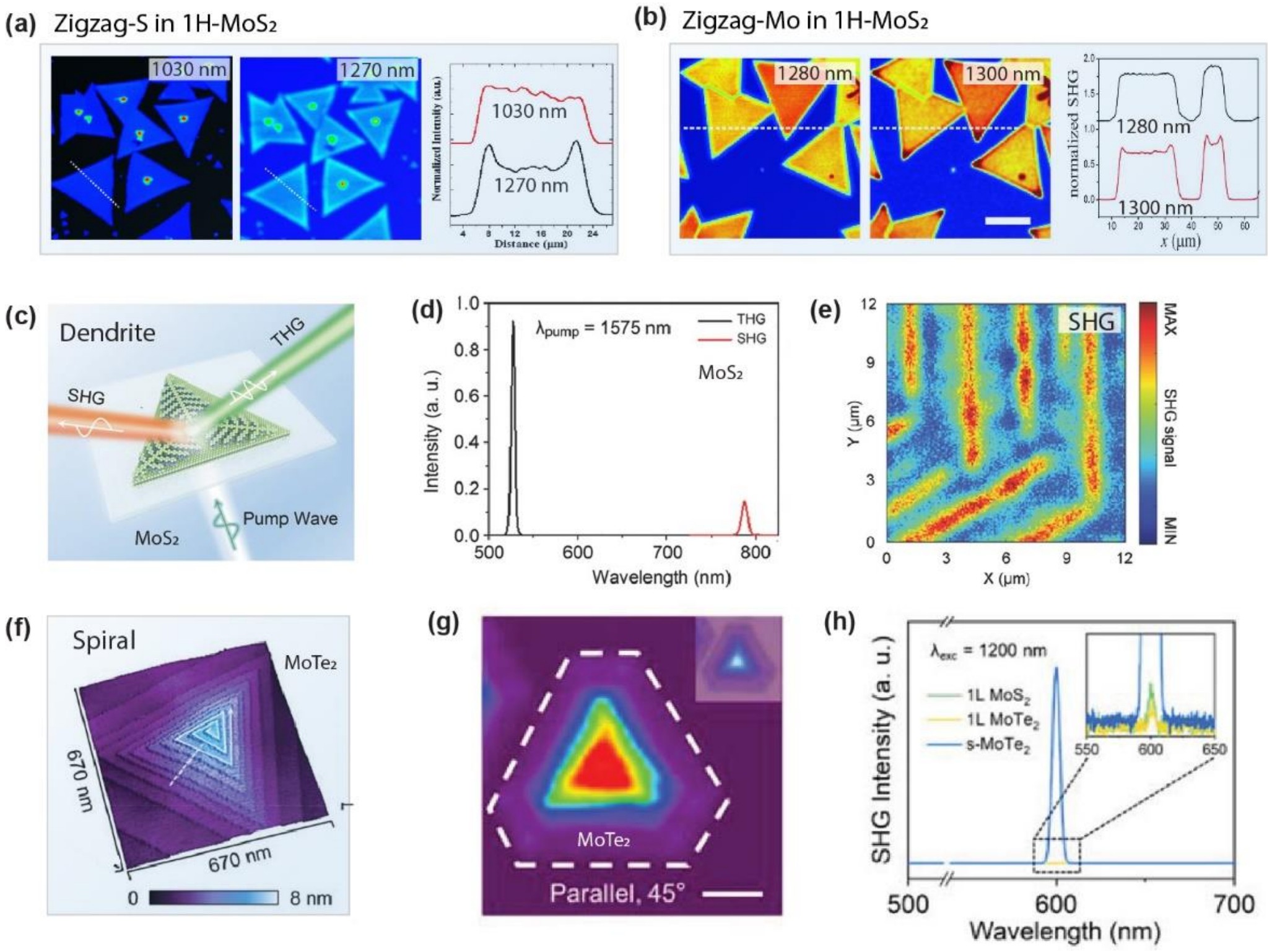

**Figure 7.** (a) SHG images of zigzag-S edged $MoS_2$ monolayer pumped at 1030 and 1270 nm, respectively; (b) SHG images of zigzag-Mo edged $MoS_2$ monolayer pumped at 1280 and 1300 nm, respectively; (c) Schematic diagram illustrating the SHG and THG response under excitation $\lambda_{pump}$ = 1575 nm; (d) SHG and THG spectra of the $MoS_2$ dendrite excited at 1575 nm; (e) Nonlinear spectral images acquired at the corresponding SHG emission wavelengths; (f) AFM image of spiral $MoTe_2$; (g) Polarization-dependent SHG mapping under parallel configuration at 45°; (h) SHG output of monolayer $MoS_2$, monolayer $MoTe_2$ and spiral $MoTe_2$ excited at 1200 nm; (a) Reproduced with permission[168]. Copyright 2018, American Chemical Society; (b) Reproduced with permission[32]. Copyright 2014, AAAS; (c-e) Reproduced with permission[62]. Copyright 2017, Wiley-VCH; (f-h) Reproduced with permission[133]. Copyright 2022, Wiley-VCH.

The manipulation of edge morphologies allows for the multidimensional engineering of the unique edge-dependent nonlinear optical properties in TMD layers. Different edge structures and stacking configurations have been observed to exhibit significant variations in various nonlinear optical processes, including second and third-harmonic generation. As TMDs approach the monolayer limit, a novel valley-specific circular dichroism emerges due to the combination of the distinctive local orbital properties of the heavy transition metal atoms and the broken inversion symmetry resulting from imbalanced charge carrier distribution.[32] When transitioning to a bilayer, SHG signals are noticeably quenched in comparison to THG signals, which being a third-order process occur regardless of layer number. Chen *et al.*[62] indicated that these distinct difference intensity peaks of SHG and THG can be observed in the dendritic $MoS_2$ domains (schematic shown in Figure 7c), when pumping at 1575 nm, the SHG and THG peaks are clearly identified (in Figure 7d). The SHG nonlinear mapping (in Figure 7e) revealed that the nonlinear property can be spatially modulated by engineering the 2D edge morphologies.

Non-centrosymmetric TMD spirals with 3D helical edge structures exhibit strong bulk second-order optical nonlinearity. Fan *et al.*[164] reported on the strong nonlinear optical effects of SHG and THG in spiral $WS_2$ structures. The contrasting SHG behavior is attributed to the broken symmetry arising from the screw structures and the exposed high-density edges. Similar enhanced SHG has also been observed in $MoS_2$ spirals by Zhang *et al.*[90] Furthermore, Te-containing TMD spirals are expected to exhibit higher nonlinear susceptibility compared to S and Se-containing TMD spirals due to the redistribution of energy level spacing caused by variations in lattice volume. Ouyang *et al.*[133] synthesized edge-rich $MoTe_2$ spirals using CVD (AFM image shown in Figure 7f) and reported the enhancement of SHG signal, as evident in polarization SHG image displayed in Figure 7g. Apart from the SHG effects resulting from the non-centrosymmetric 3R-like stacking configuration, significant near-field enhancement is observed in spirals. Simulations using the discrete dipole approximation predict this enhancement on layer edges of a 3R stacked pyramid, with the near-field strength decreasing from the bottom to the top of the pyramid but remaining continuous along the step edges.[133] The similarity of TMD spiral structures suggests a similar near-field enhancement along the spiral lines. The observed enhancement in SHG in spirals, compared to the SHG intensities of monolayers of $MoS_2$ and $MoTe_2$, can be attributed to the synergistic effect of the non-centrosymmetric stacking and the continuous near-field enhancement of the edge states along the spiral lines. Furthermore, the tunability between different TMD spiral morphologies, including hexagonal, triangular, and mixed shapes, enables the versatile engineering of 3D spiral edge structures, resulting in the modulation of SHG properties in 3D space.[132] These findings highlight the potential of TMD spirals as platforms for tailoring and engineering nonlinear optical effects, opening up new avenues for advanced photonic applications.

### 4.3. Catalysis

The hydrogen evolution reaction (HER) is a crucial process for generating clean and sustainable energy. To be an effective HER catalyst, it is essential to have a suitable free energy of hydrogen adsorption ($\Delta G_{H^*}$) that closely matches that of the reactant or product.[28, 29, 169, 170] In particular, if $\Delta G_{H^*}$ is too high (indicating weak binding), the reaction rate is limited by the adsorption of hydrogen atoms. Conversely, if $\Delta G_{H^*}$ is too negative (indicating strong binding), the reaction becomes hydrogen-desorption limited. Thus, finding the optimal $\Delta G_{H^*}$ is essential for developing efficient HER catalysts.

2D atomic TMD layers have emerged as highly promising catalysts for HER due to their versatile chemistry and the ability to employ various physical, chemical, and electronic engineering strategies. Numerous review papers have extensively summarized these strategies, including phase and defect engineering, chemical doping, and the formation of hybrid structures.[28, 29, 171, 172] Among these approaches, edge engineering in TMD layers has proven to be an effective method for modulating the HER properties.

One key advantage of edge engineering is that the metallic edges of TMDs exhibit adsorption free energy of hydrogen close to that of platinum and thermoneutral conditions. For instance, the $\Delta G_{H^*}$ value of $MoS_2$ zigzag edges is approximately 0.08 eV. In Figure 8a,b, Lin et al[29]. summarized $\Delta G_{H^*}$ values at different edges of various TMDs against the absorption free energy of H-X (where X represents the chalcogen). When $\Delta GH_{X^*} > 0$, the reaction tends to form $H_2X$ while desorbing the H-X group. Conversely, reducing the $\Delta G_{HX^*}$ values enhance stability. It revealed that besides $MoS_2$, several other TMDs, such as $MoSe_2$, $WS_2$, $WSe_2$, and $TaS_2$, also exhibit high promise for HER with their 2H or 1T edges. It is worth noting that substituting Se with S in TMDs can slightly adjust the $\Delta G_{H^*}$ value towards a more thermoneutral state. However, this substitution also leads to a reduction in stability.

The HER performance is influenced by the catalytic activity and number of active sites. While CVD-grown samples may have a limited number of active sites compared to other fabrication methods, the flexible control of atomic edge configurations, edge morphologies, and edge atomic density in CVD-grown TMDs provides a platform to investigate the fundamental mechanisms associated with different atomic edge configurations, morphologies, and edge densities. This can be achieved through the combination of in situ electrochemical techniques such as scanning electrochemical cell microscopy (SECCM) and in-situ electrochemical microdevice testing.

SECCM is a pipette-based tip technique that allows for imaging surfaces with nanometer-scale lateral resolution, as shown in Figure 8c. It can be used to conduct localized electrochemical measurements and imaging of electrochemical activity.[173] [174] Takahashi *et al.*[175] used SECCM to image and quantitatively analyse the catalytically active sites of CVD-grown monolayer $MoS_2$ edges. The SECCM image in Figure 8d revealed a higher current response at the edges compared to the basal plane region, indicating the edges addressed more active HER activity. Additionally, polarization curves in Figure 8e indicated a lower overpotential at the edges of the $MoS_2$ monolayer compared to the basal plane. This further supports that the edge atoms are more active than basal-plane atoms in HER catalysis.

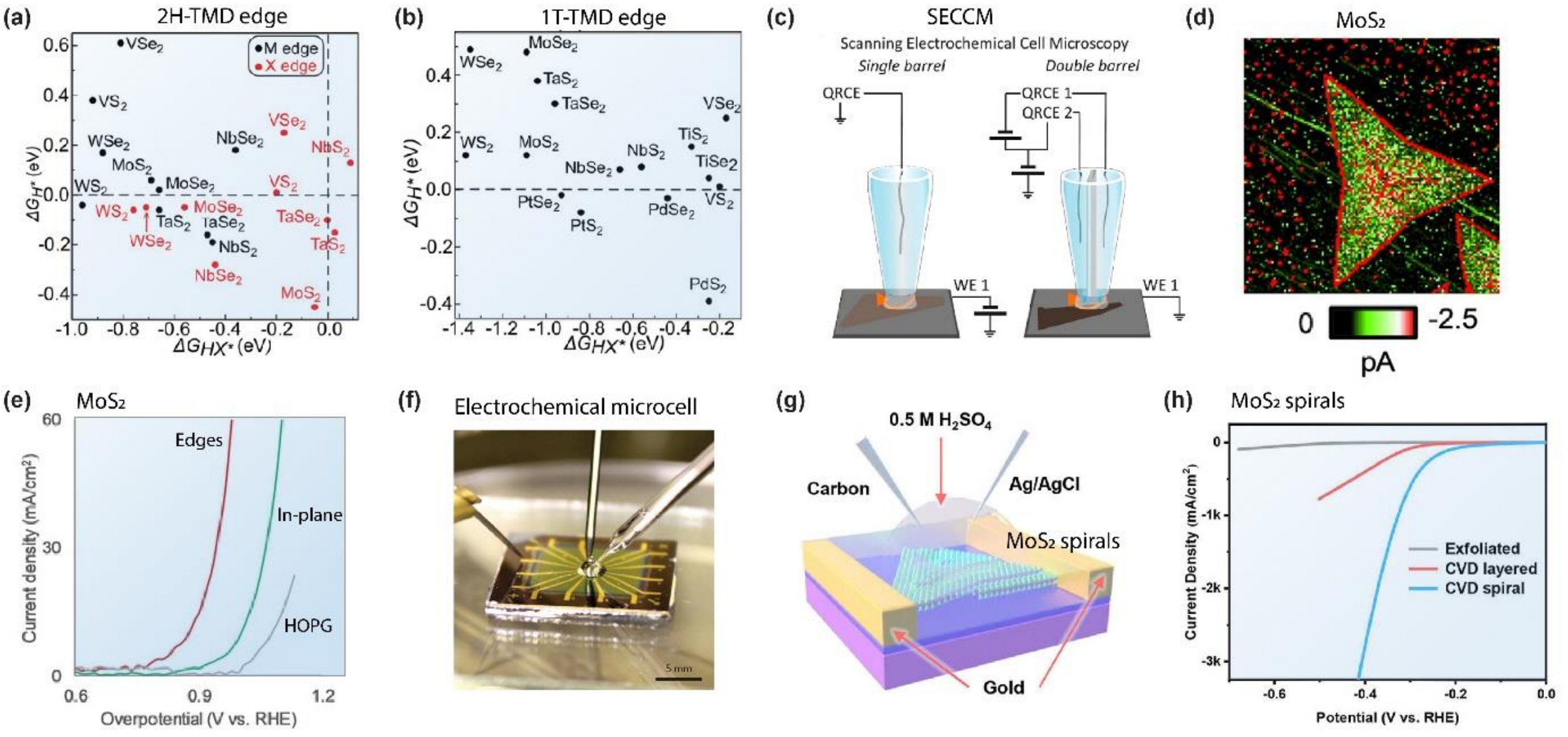

**Figure 8.** (a) Plots of the $H_2$ adsorption free energy ($\Delta G_{H^*}$) as a function of the HX adsorption energy ($\Delta G_{HX^*}$, X=S or Se) at 2H-TMD edges; (b) Plots of the $\Delta G_{H^*}$) as a function of the $\Delta G_{HX^*}$, at 1T-TMD edges; (c) The schematic of scanning electrochemical cell microscopy technique (SECCM); (d) The SECCM current images of $MoS_2$ monolayer on a HOPG substrate; (e) Polarization curves on the $MoS_2$ edges, basal plane and HOPG edge regions; (f) Photograph of the electrochemical microcell; (g) Schematic illustration of an electrochemical microcell setup for spiral $MoS_2$; (h) Polarization curves of multi-layered and spiral $MoS_2$ flakes layered and spiral $MoS_2$; (a, b) Reproduced with permission[29]. Copyright 2020, Wiley VCH; (c) Reproduced with permission[173]. Copyright 2023, American Chemical Society; (d, e) Reproduced with permission[175]. Copyright 2019, Wiley-VCH; (f) Reproduced with permission[175]. Copyright 2019, Wiley-VCH; (g, h) Reproduced with permission[176]. Copyright 2022, Wiley-VCH

In addition, microelectrochemical cell has been developed for measuring the localized catalytic activity of TMD layers.[177, 178] Through electron beam lithography patterning, it is possible to selectively measure the catalytic properties of either the basal plane or the edges. The individual TMD structures were investigated using a three-electrode configuration, with glassy carbon as the counter-electrode, an Ag/AgCl electrode as the reference electrode, and a gold pad contacting the TMD layers as the working electrode. To ensure that the measured activity was solely due to the TMD testing region, the area outside the target region was covered with cured poly(methyl methacrylate) (PMMA) resist.

This microreactor setup enabled position-resolved measurement of the HER activity of TMD edges versus the basal plane. By selectively exposing the edges and basal plane of 2H-$MoS_2$ and different edges of 1T-$WTe_2$, Zhou *et al.*[179] reported the (100) edge of $WTe_2$ exhibited the highest catalytic activity among the tested atomic edge configurations. Moreover, the impact of edge morphologies on catalytic activity can also be investigated using the microreactor setup. Tong *et al.*[176] designed a microreactor setup on a spiral $MoTe_2$ domain and found that the regions near the edge step of the spiral structures, which easily form Te vacancies, exhibited lower $\Delta G_{H*}$ and served as additional active sites. An on-chip microcell based on a single spiral $MoTe_2$ achieved an ultrahigh current density of 3000 mA/cm$^2$ at an overpotential of 0.4 V, surpassing the performance of exfoliated counterparts and CVD-grown layered $MoTe_2$. The atomically layered continuous spiral pyramid structure provided abundant edges as highly active sites for HER. Furthermore, Su *et al.*[180] discovered that the helical edges in spiral $MoS_2$ can facilitate the formation of micro helical currents, leading to magnetic heating and boosting the electrocatalytic activity. This finding highlights the potential for harnessing spiral structures in TMD layers to enhance HER performance.

### 4.4. Others

Surface-induced polymer crystallization, involving heterogeneous nucleation and crystallization, is a significant area of research in polymer science.[181-183] Understanding the interaction between atomic motifs, such as edges, grain boundaries, and crystalline surfaces, with polymer chains poses a challenge due to the complexity of these interactions. However, atomic TMD layers offer an ideal platform for studying the interaction between atomic edges and polyethylene chains in dilute solution assembly processes, due to their well-isolated edges and controlled atomic arrangement. Zhou *et al.*[184] demonstrated that zigzag-Mo edges act as preferred nucleation sites and strongly interact with polyethylene chains. The chains align parallel to the edges, resulting in arrays of lamellae perpendicular to the edges (Figure 9a-c). The result also revealed that the atomic edge configurations play a crucial role, influencing the crystallization outcomes. These findings indicate that the atomic edge structures of TMDs can significantly impact their interactions with n-alkane chains.

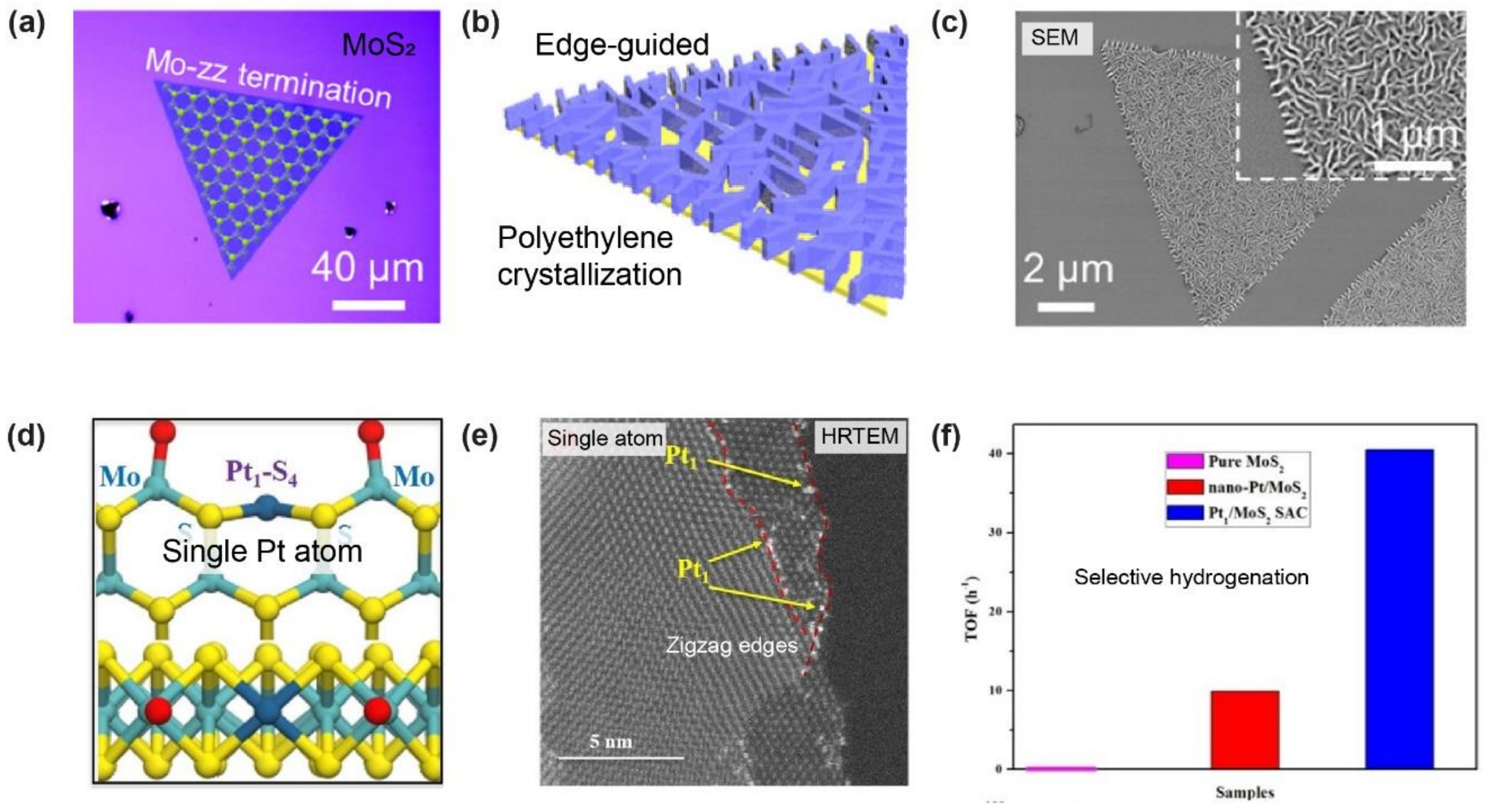

**Figure 9.** (a-c) zigzag-Mo edge-guide polyethylene crystallization; (d-f) zigzag edge-confined single Pt atoms for excellent catalytic activity and selectivity for the hydrogenation. (a) Reproduced with permission[184]. Copyright 2022, American Chemical Society. (b) Reproduced with permission[185]. Copyright 2021, European Chemical Societies Publishing.

Atom catalysis has gained considerable attention due to its potential for enhancing the efficiency and sustainability of chemical reactions.[186-189] Single atom catalysts, known for their high activity, selectivity, and stability, hold promise for various applications.[190] CVD-grown TMDs, with their rich edge structures, provide an excellent platform for anchoring single atoms, leading to improved catalytic performance.[191-194] Lou *et al.*[185] designed 2D $MoS_2$ zigzag edge-confined single Pt atom catalysts with a unique HO−Mo−S−Pt1−S−Mo−OH configuration (Figure 9d-f). The Pt atoms in these catalysts possess distinctive electronic structures resulting from the Pt1-S4 configuration, creating an enzyme pocket-like active site that restricts the adsorption of crotonaldehyde through steric hindrance effects, thereby achieving high selectivity.

## 5. Summary and Outlook

TMD atomic layers have garnered significant attention, due to their diverse properties and potential applications. Besides tailoring the properties via chemical compositions, phase structures (2H, 1T, and 1T'), the presence of various edge configurations, variable edge density, and edge morphologies in 1D, 2D, and 3D forms also show great promise. This comprehensive review has attempted to shed light on the current state of research in TMD edge engineering, focusing on growth exploiting the versatility of CVD to manipulate atomic edge configurations and achieve multi-dimensional morphologies, and on the vast range of unique properties associated with these edge configurations and structures.

Versatile growth control strategies have been reported for TMD edge engineering. These include the metal particle driven growth mechanism, space confinement techniques, substrate surface step engineering, and surface chemical modification. Each approach offers specific advantages in controlling the edges of 1D nanoribbons, 2D nanoribbons, and compact shapes, as well as the growth of 3D spirals and nanotubes. Notably, the space confinement strategy utilizing carbon or BN nanotubes for 1D nanoribbons or 3D nanotubes achieves exceptionally high atomic edge densities. However, challenges related to low production yields resulting from diffusion limitations or surface strain hinder the scalability of this method.

On the other hand, substrate surface engineering strategies such as tailored surface steps and chemical modifications hold significant promise for synthesizing large-scale, high-density 1D nanoribbon edges. Precise control of growth enables the attainment of controllable edge densities up to $6.0 \times 10^6$ $\mu m^{-2}$ with excellent uniformity. Furthermore, the screw dislocation-driven growth of 3D spirals under low supersaturation conditions leads to the creation of unique continuous helical edges with ultrahigh atomic edge densities. The introduction of chirality in these 3D edges presents exciting opportunities for various unique properties.

Despite the progress made, several challenges remain. The relationship between edge density and material properties necessitates the development of techniques for accurately detecting and quantitatively calculating atomic edge densities. The integration of image recognition techniques with artificial intelligence could be useful for addressing this challenge. Additionally, improving the reproducibility of edge control and achieving wafer-scale, unidirectional orientation of edges are crucial objectives. New strategies focusing on uniformity control and providing a stable growth environment for various edge configurations, such as nanoribbon and spiral growth techniques, are also of great importance.

Beyond the structural modulation of various TMD edge morphologies and atomic edge types, future research aimed at exploring atomic precision engineering of TMD edges and investigating their interlayer interaction with the underlying substrates holds great promise for unlocking edge related-functionalities.

Advanced edge engineering may be possible through the combination of CVD with other fabrication techniques. For instance, integrating CVD growth of 1D nanoribbons and 2D dendrites or compacts with the rolling-up technique enables the creation of 3D nanotubes with rich edge structures. This approach allows for precise control over the chiral degree, as the atomic edge configuration and structure features are provided by CVD, while the subsequent rolling-up process shapes the 1D or 2D edges into 3D nanotubes.

The versatile edge structures of TMDs hold significant potential for traditional electronics, optical and catalytic uses as well as for quantum applications and other emerging technologies. Continued research and development in TMD edge engineering will pave the way for exciting advancements in electronics, optoelectronics, catalysis, and other fields.

**Acknowledgements**

We acknowledge the funding support from Agency for Science, Technology and Research (#21709). K.E.J.G. acknowledges support from a Singapore National Research Foundation Grant (CRP21-2018-0001).